# Reliability Assessment of Microgrid with Renewable Generation and Prioritized Loads

Osama Aslam Ansari, Nima Safari, and C. Y. Chung
Smart Grid and Renewable Energy Technology (SMART) Lab
Department of Electrical and Computer Engineering
University of Saskatchewan
Saskatoon, Saskatchewan, Canada
{oa.ansari, n.safari, c.y.chung}@usask.ca

*Abstract*— **With the increase in awareness about the climate change, there has been a tremendous shift towards utilizing renewable energy sources (RES). In this regard, smart grid technologies have been presented to facilitate higher penetration of RES. Microgrids are the key components of the smart grids. Microgrids allow integration of various distributed energy resources (DER) such as the distributed generation (DGs) and energy storage systems (ESSs) into the distribution system and hence remove or delay the need for distribution expansion. One of the crucial requirements for utilities is to ensure that the system reliability is maintained with the inclusion of microgrid topology. Therefore, this paper evaluates the reliability of a microgrid containing prioritized loads and distributed RES through a hybrid analytical-simulation method. The stochasticity of RES introduces complexity to the reliability evaluation. The method takes into account the variability of RES through Monte-Carlo state sampling simulation. The results indicate the reliability enhancement of the overall system in the presence of the microgrid topology. In particular, the highest priority load has the largest improvement in the reliability indices. Furthermore, sensitivity analysis is performed to understand the effects of the failure of microgrid islanding in the case of a fault in the upstream network.**

## I. Introduction

For the last few decades, there has been an increasing concern about the climate change and depletion of non-renewable energy sources. Increased public awareness has called for reduction of carbon emissions which are one of the main sources of climate change. This requires gradual yet steady replacement of conventional coal-based power plants with environmental-friendly renewable energy sources (RES). Smart grid technologies have been developed to facilitate large-scale integration of RES and to tackle different challenges associated with it. Microgrids are one of the main building blocks of smart grids [1]. Microgrids do not only offer the integration of RES in the distribution system but also provide attractive features such as islanded mode of operation and higher flexibility. In order to fully understand the cost-benefit analysis of distribution system containing microgrids, it is important to take into account the reliability of the electric supply at the customers' ends. Maintaining system reliability is one of the primary motives of any utility. Significant costs and penalties are incurred as a result of system interruptions. Hence not only the company's reputation but also the financial reasons force utilities to ensure an acceptable level of system reliability. In this regard, several standards have been developed to make sure that utilities guarantee the reliable supply of electricity to their customers. In such scenarios, reliability studies become crucial for utilities. Reliability studies can also be considered in distributed energy resources (DER) sizing, siting and operation, and reinforcement of the crucial elements in distribution network [2], [3], [4].

The inclusion of RES introduces complexity in their modeling for reliability studies. Furthermore, the output of distributed generation (DGs) based on RES are energy limited and sporadic in nature. In this case, the priority order of the loads should be considered to maintain electric supply to the most sensitive loads in the event of insufficient generation. In literature, several techniques have been presented to evaluate the reliability at customer load points in the microgrid and of the microgrid as a whole. They can be categorized into analytical methods, simulation methods and hybrid methods. In [5], a method based on Monte-Carlo simulation to evaluate reliability of an active distribution system with multiple microgrids is proposed. It has been shown that the inclusion of DGs and storage increases the overall reliability of the system. The method discretizes the output of RES and evaluates probability for each step. In [6] and [7], the reliability of a microgrid in islanded mode is evaluated. In [6], Monte-Carlo simulation is used to model component failure and component repair and historical data for RES. In [7], fault tree analysis is adopted which can become mathematically involved for large systems. Since grid-connected mode of microgrid is not taken into account in both of the papers, the values for the reliability metrics are different from the actual values. Markov modeling is used to evaluate the reliability of microgrid with photovoltaic (PV) generation and energy storage systems (ESSs) in [8]. However the proposed method uses a simple two-state model for PV generation which is insufficient to incorporate the highly intermittent nature of PV and other RES such as wind energy. In [9], the intermittent nature of PV based DGs is ignored by considering that the ESSs are sufficiently sized to make DGs dispatchable. The values of reliability indices can be markedly different if intermittency is taken into account. Some papers such as [10] evaluated the reliability indices of a small isolated power systems with RES. However, they do not take into account the abilities of a microgrid such as its different modes of operation. . In [11], a hybrid model is presented to find out the reliability of a microgrid in the presence of renewable DGs and ESSs. The method does not consider the priority order of the loads. Also in one of the cases, the method assumes that battery helps intermittent DGs to supply the entire load or a large part of the load.

This paper presents a hybrid method comprising of both analytical and simulation techniques to evaluate the reliability at customers' load points in a microgrid containing wind and solar power generation and prioritized loads. State sampling Monte-Carlo simulation is combined with the analytical method. Simulation methods take into account the chronology of variation of RES and the priority order of the loads whereas analytical methods consider the topology of the network and the failure and repair rates of the networks' components.

The paper is formatted as follows. Section II briefly introduces the microgrid. Section III presents the modeling of wind and PV generation sources for the reliability evaluation. Section IV delves into the proposed method for calculating the reliability indices of the microgrid. The results are indicated in Section V. Section VI provides the conclusion of the work.

## II. Microgrids

A microgrid is a group of interconnected loads, DERs (such as DGs and ESSs), and management systems that can operate either in grid connected mode or islanded mode (Fig. 1). From the utility's point of view, a microgrid appears as a single controllable entity that can consume or supply power depending upon the total generation and the total load inside the microgrid.

The ability of a microgrid to operate in islanded mode increases the reliability of the load points. The islanding mode can occur if there is a fault within the microgrid or in the upstream network to which it is connected. In islanded mode of operation, the loads depend upon the power generation of DGs connected in the microgrid. In the case when DGs are unable to supply all of the load, energy management systems can voluntarily curtail the non-sensitive loads utilizing advanced switches and control strategies. This ensures that the sensitive loads are not interrupted [1], [4].

## III. Modeling Renewable Energy Sources

One of the key features of the microgrid topology is its ability to integrate DERs such as renewable DGs, dispatchable DGs, ESSs and electric vehicles (EVs) in a seamless manner. DERs not only provide an alternative to the generation expansion planning but also assist in improving the reliability of the distribution system to which they are connected [4], [12], [13]. Nevertheless, RES poses different challenges to the operation and planning of the power system. The most significant of these is their intermittent behavior. The output of RES depends on various factors including weather conditions and geographical location. Moreover, RES have widely different characteristics based on their source of energy, for instance, wind and solar. Hence it is difficult to use a general model for all RES or to use the same models as utilized for conventional generators. RES generators also have their own failure rates and repair time. It has been observed that the unavailability of RES is primarily because of the unavailability of their energy source rather than the failure of RES generators [14]. The method presented in this work uses the above observation.

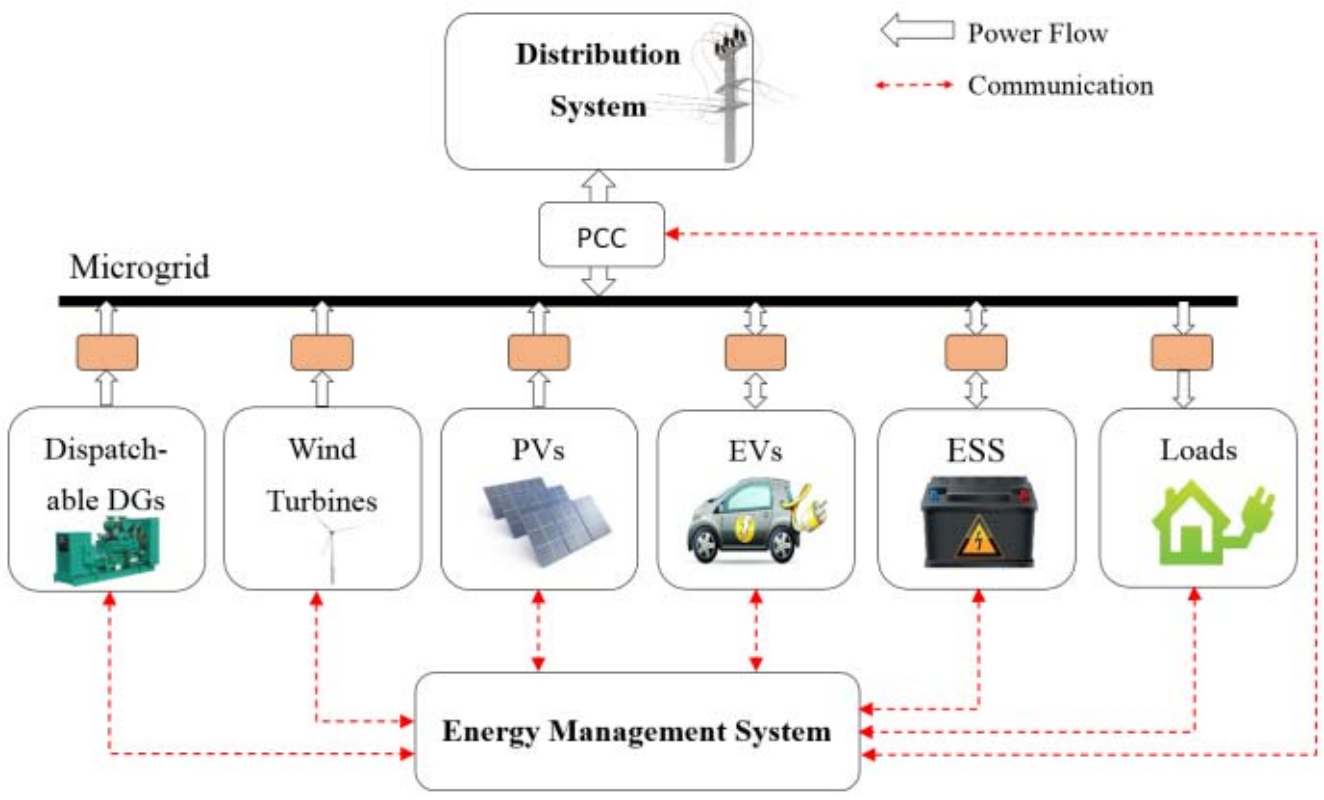

Fig. 1. Typical structure of a microgrid

In reliability studies involving simulation techniques, it is necessary to generate synthetic data for RES power generation. The next parts of this section introduces the models to generate the data for this purpose.

### A. Wind Power Generation Modeling

Wind energy is the most developed and the fastest growing form of renewable energy. Wind power depends upon different factors including wind speed, wind direction, and geographical locations etc. However, wind power is largely dependent on the wind speed. Therefore, in this paper, the wind power is modeled using the wind speed.

Different distributions such as normal [15] and Weibull [16] are used to model the wind speed. In general, two parameter Weibull distribution provides a better fit for modeling the wind speed [17]. In this paper, the model for wind speed distribution function is obtained from [16] which uses Weibull distribution.

The cumulative probability distribution for the wind speed $v$, $F_w(v)$ is given as:

$$F_w(v) = 1 - e^{-(v/c)^k} \tag{1}$$

where, $c$ is the scale parameter and $k$ is the shape parameter. The values for two parameters are obtained using historical data. They are adopted from [16] and given in Table I.

In order to simulate the wind speed values for a required duration, the inverse transformation method is implemented [18]. Inverse transformation method states that if a random variable (in this case $v$) follows the U [0, 1] uniform distribution, then the random variable $X = F^{-1}(U)$ has a continuous cumulative probability distribution function $F(X)$. Using this principle, (1) can be expressed as:

$$v = c[-\ln(1-X)]^{1/k} \tag{2}$$

where $X$ is a random number between 0 and 1. Since $X$ is a random number, (2) can be also be expressed as:

$$v = c[-\ln(X)]^{1/k} \tag{3}$$

TABLE I
WEIBULL DISTRIBUTION DATA

| Parameter | Region 1 | Region 2 |
|---|---|---|
| Mean speed (m/s) | 7.0 | 7.5 |
| *k* | 2.62 | 3.18 |
| *c* | 7.88 | 8.46 |

Using (3), daily values of wind speed can be simulated by generating random number X for each day.

The wind power is obtained from the wind speed by using the WTG power curve shown in Fig. 2. The power curve expresses the wind power as the function of wind speed and is provided by the wind turbine manufacturer. The power curve of WTG is usually characterized by the following parameters:
Rated speed ($v_{\text{rated}}$): The speed at which the maximum power can be extracted from the WTG.
Rated power ($P_{\text{rated}}$): The maximum power that can be produced by WTG at the rated wind speed.
Cut-in speed ($v_{\text{cut-in}}$): The minimum wind speed required to produce the power from WTG.
Cut-out speed ($v_{\text{cut-out}}$): The maximum speed at which WTG can operate. Usually after this wind speed, the WTG is shut down due to safety reasons.

The wind power curve is given as:

$$P_{WTG} = \begin{cases} 0 & 0 < v \le v_{\text{cut-in}} \\ av^3 - bP_{\text{rated}} & v_{\text{cut-in}} < v \le v_{\text{rated}} \\ P_{\text{rated}} & v_{\text{rated}} < v \le v_{\text{cut-out}} \\ 0 & v_{\text{cut-out}} \le v \end{cases} \tag{4}$$

where, a and b are given as:

$$\text{a} = \frac{P_{\text{rated}}}{v_{\text{rated}}^3 - v_{\text{cut-in}}^3} \qquad \text{b} = \frac{v_{\text{cut-in}}^3}{v_{\text{rated}}^3 - v_{\text{cut-in}}^3} \tag{5}$$

Using the wind power curve, the simulated wind power is obtained from the wind speed. The simulated wind speed and wind power for one of the two WTGs is shown in Fig. 3. Table II provides the data for the two wind turbines.

### B. Solar Power Generation Modeling

With the development of new solar panels, the penetration of solar power is increasing and further growth is expected in the near future. Solar power of the solar panels mainly depends upon the solar radiation and temperature. The temperature dependency of solar power is non-linear and introduces complexity to the modeling. In this work, solar radiation is utilized to obtain the output power of solar panels. Historical data of solar radiation is used to find out the probability distribution of solar radiation. The historical data for 5 years is obtained from [19]. It has been shown that the beta distribution fits more accurately to solar radiation as compared to gamma and logarithmic distribution [20]. Therefore in this paper, beta distribution is fitted on the historical data of solar radiation.

TABLE II
WIND TURBINES DATA

| Parameter | Region 1 | Region 2 |
|---|---|---|
| Rated Power (kW) | 2000 | 1500 |
| Rated speed (m/s) | 15 | 12 |
| Cut-in speed (m/s) | 3 | 3 |
| Cut-out speed (m/s) | 25 | 25 |

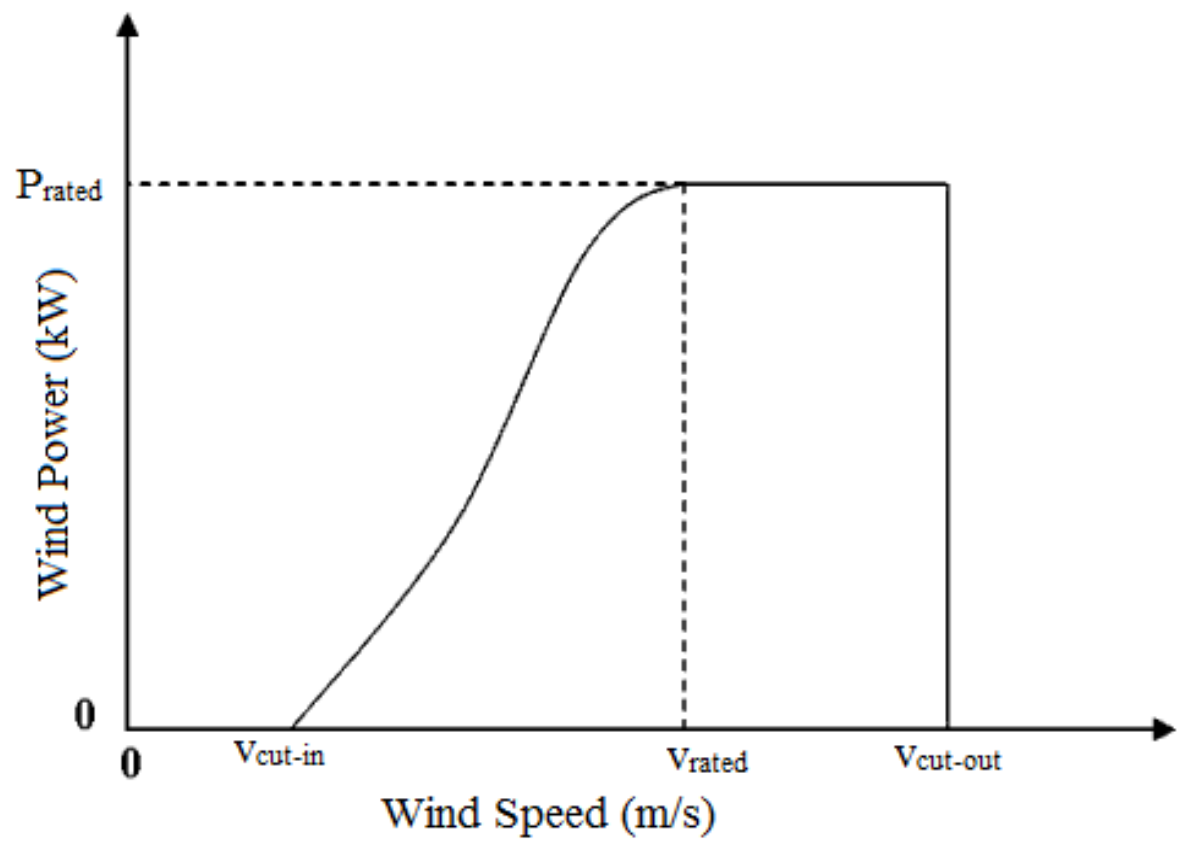

Fig. 2. Wind power curve.

The probability distribution function of beta distribution is given as:

$$f(x) = \frac{x^{\alpha-1}(1-x)^{\beta-1}}{\text{B}(\alpha,\beta)} \tag{6}$$

where, B in terms of gamma function (Γ) is defined as:

$$\text{B}(\alpha,\beta) = \frac{\Gamma(\alpha)\Gamma(\beta)}{\Gamma(\alpha+\beta)} \tag{7}$$

After fitting the beta distribution on historical data, the two parameters for beta distribution obtained from statistics and machine learning toolbox of MATLAB are as follows:

$$\alpha = 1.03745 \qquad \beta = 1.38279 \tag{8}$$

After obtaining the corresponding probability distribution, inverse transformation method is applied. Using the inverse transformation method, solar radiation is predicted for the required interval of time. Similar to wind power curve, the solar power curve expresses solar power in terms of solar radiation. The expression for output power of PV in terms of solar radiation is given as [14]:

$$P_{\text{PV}} = \begin{cases} P_{\text{sn}} \dfrac{G^2}{G_{\text{std}} R_{\text{c}}} & 0 \le G < R_{\text{c}} \\ P_{\text{sn}} \dfrac{G}{G_{\text{std}}} & R_c \le G \le G_{\text{std}} \\ P_{\text{sn}} & G > G_{\text{std}} \end{cases} \tag{9}$$

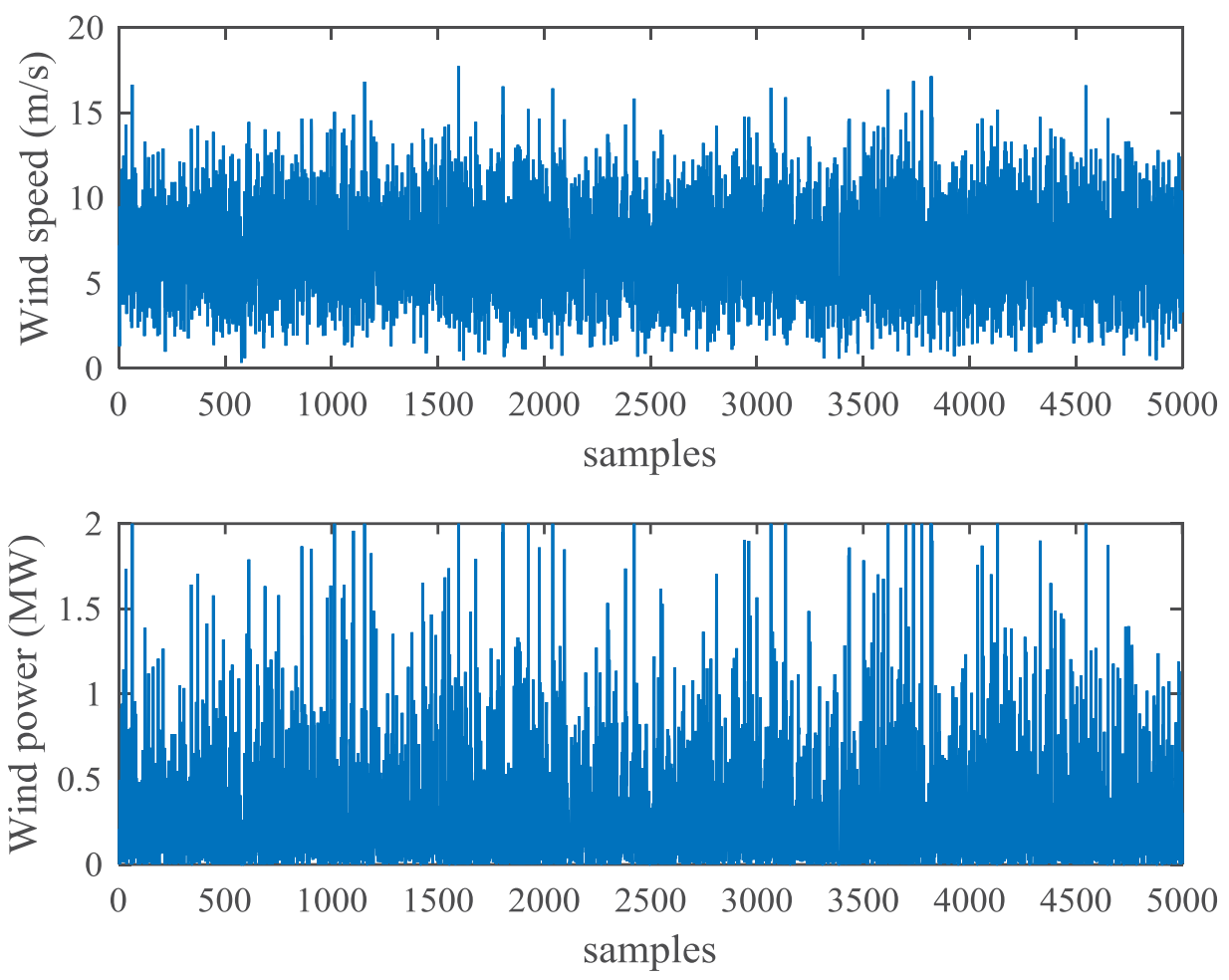

Fig. 3. Simulated wind speed and wind power for WT A where, $P_{PV}$ is solar output power in MW. $G$ is simulated solar radiation in W/m$^2$. $G_{std}$ is solar radiation in the standard environment. Usually this value is set to 1000 W/m$^2$. $R_c$ is a certain radiation point set usually to 150 W/m$^2$. $P_{sn}$ is the rated output power of solar panels.

## IV. Reliability Evaluation Method

### A. Reliability Indices

For evaluating the reliability of customer load points in a distribution system, the system average interruption frequency index (SAIFI), the system average interruption duration index (SAIDI), the customer average interruption duration index (CAIDI), the energy not supplied (ENS) and average energy not supplied (AENS) indices are used. These are defined as [21]:

$$\text{SAIFI} = \frac{\sum \lambda_i N_i}{N_T} \tag{10}$$

$$\text{SAIDI} = \frac{\sum U_i N_i}{N_T} \tag{11}$$

$$\text{CAIDI} = \frac{\text{SAIDI}}{\text{SAIFI}} \tag{12}$$

$$\text{ENS} = \sum_{i \in LP} U_i L_i \tag{13}$$

$$\text{AENS} = \frac{\text{ENS}}{N_T} \tag{14}$$

where, $\lambda_i$ is the failure rate, $U_i$ is the outage time, $N_i$ is the number of customers, and $L_i$ is the total load at load point $i$. $N_T$ is the total number of customers in the system.

### B. Evaluation Method

The steps for reliability evaluation are as follows:

**Step 1**. Obtain the cumulative distribution functions of wind and solar radiation from their historical data.

**Step 2**. Use cumulative distribution functions, and the inverse transformation method to generate the simulated values of wind speed and solar radiation for required number of samples. This requires generation of random number $X$ for each sample. Here one sample can be considered as one day.

**Step 3**. Convert the wind speed to wind power using wind power curve (4) – (5) and solar radiation to solar power using (9).

**Step 4**. At each day, sample the output of all WTGs and PV panels. The combined output of all RES is compared with the load. In the case if RES cannot supply all of the loads, then the load with highest priority is supplied first followed by the next load in the priority list and so on.

**Step 5**. Step 4 is performed for each day of the year. The number of occurrences at which the load is supplied by RES for each of the load points is counted. At the end of year, the probability of RES supplying different load points is calculated. Let $P^i_{\text{RES}}$ be the probability that RES can supply the load at load point $i$. This probability value along with the system stochastic data (failure rate and repair time) are used to obtain the average failure rate, unavailability and repair time at each load point at the end of the year using the following equation:

$$\begin{aligned} \lambda_i &= \sum \lambda_j + (1 - P^i_{\text{RES}})\lambda_{\text{up}} \\ U_i &= \sum \lambda_j r_j + (1 - P^i_{\text{RES}})\lambda_{\text{up}} r_{\text{up}} \end{aligned} \tag{15}$$

where, $\lambda_i$ and $U_i$ are the failure rate and unavailability at load point $i$, respectively, $\lambda_{\text{up}}$ and $r_{\text{up}}$ are the failure rate and repair time of upstream network, respectively, and $\lambda_j$ and $r_j$ are the failure rate and repair time of microgrid network components, respectively that results in interruption at load point $i$.

**Step 6**. Afterwards, the system indices SAIFI, SAIDI, CAIDI, ENS and AENS are calculated using (10) – (14).

**Step 7**. Steps 4-6 are repeated for each year until the variation in system indices is less than the specified tolerance or maximum number of iterations is reached.

In (15), each expression consists of two terms. The first term arises from the stochastic data of the network components. The second term is derived from the following conditional probability equation.

$$\begin{aligned} \lambda &= (\lambda \mid \text{RES can supply}) \times P^i_{RES} \\ &\quad + (\lambda \mid \text{RES cannot supply}) \times (1 - P^i_{RES}) \end{aligned} \tag{16}$$

In the first part of above expression, in the case of failure in the upstream network, if RES can supply the load point $i$, the failure in the upstream network will not result in any interruption at that load point. Hence, the (16) reduces to:

$$\lambda = (1 - P^i_{\text{RES}}) \times \lambda_{\text{up}} \tag{17}$$

The $P^i_{\text{RES}}$ calculated through simulation takes into account the intermittency and variability of RES. During a sample, if RES generation is greater than the load, the excess generation would be curtailed.

## V. Case Study and Results

### A. Test System

The proposed method is applied to a modified version of the distribution system connected to bus 5 of the Roy Billinton

TABLE III
LOAD DATA

| Load Point | Load Level (kW) | Type of Load | No. of Customers |
|---|---|---|---|
| LP2 | 1000 | Commercial | 100 |
| LP3 | 3000 | Office loads | 300 |
| LP4 | 1000 | House loads | 250 |
| LP9 | 500 | Governmental | 50 |

Test System (RBTS) [21]. The advantages of using the RBTS include the availability of stochastic data for network components, and manageable system size.

The test system is shown in Fig. 4. Table III provides the assumed load data for the system. For this system, in the case of a fault in a feeder section, assume 4 for an instance, the management system acts so as to reduce interruption duration for the customers. In this case, firstly F1 would open and then isolating switches connected to section 4 would open to isolate the faulty section from the rest of the network. Then F1 and normally open (N/O) switch would close to restore the supply to rest of the customers. In this situation, LP3 and LP4 would be interrupted until section 4 is repaired completely whereas, the other load points would be interrupted for the duration equal to the switching time of F1, N/O and isolating switches. As mentioned earlier, a priority list is constructed to serve the most sensitive load first in the case of islanded mode of operation resulting from failure in the upstream network. In this case it is assumed that government load (LP9) has the highest priority followed by office loads (LP3) and then house loads (LP4). On the other hand, commercial loads (LP2) are assumed to have the lowest priority.

Following four cases are studied to consider the effects of DGs on reliability of microgrid.

**Case 1**: Without DGs.
**Case 2**: With 4 WTGs as DGs.
**Case 3**: With 2 WTGs and 2 PV panels as DGs.
**Case 4**: Case 3 with variable load.

The ratings for the WTG and PV panels are given in table IV. These ratings are selected considering that the capacity factors of DGs based on RES are quite low. Hence the combined rated power of all DGs is higher than the total load in the microgrid.

The failure rates and repair times of the test system are given in [21]. The failure rate and the repair time of upstream network are assumed to be 0.5 f/yr and 10 hours, respectively. The switching time and repair time are assumed to be 3.5 hours and 30 hours, respectively. These assumed values are frequently used in the literature. The maximum number of samples is set to 100,000. The variable load model is obtained from [21].

TABLE IV
RENEWABLE DGS DATA

| Type | Location | Rated Power (kW) |
|---|---|---|
| WTG 1 | LP7 | 2000 |
| WTG 2 | LP10 | 1500 |
| PV 1 | LP1 | 2000 |
| PV 2 | LP8 | 2000 |

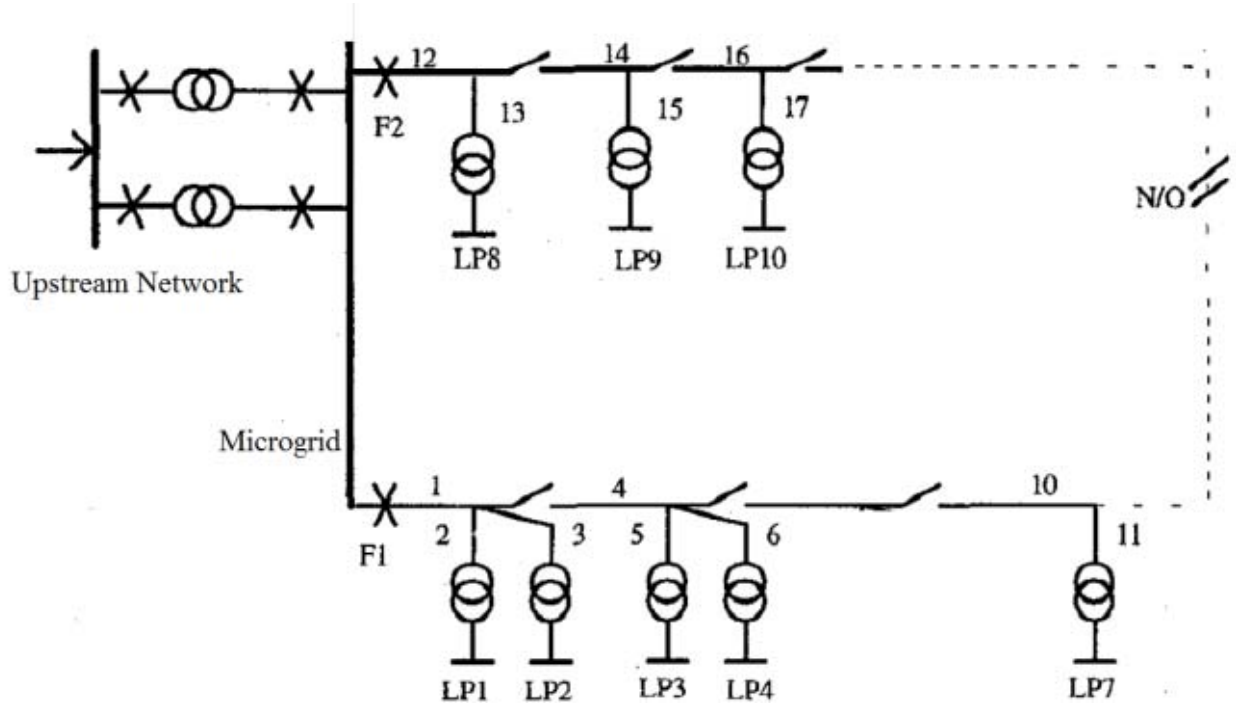

Fig. 4. Modified Bus 5 of RBTS [21]

## B. Results

The application of the method on four cases indicates that the presence of DGs in the microgrid increases the overall reliability of the system. Although the values of SAIFI do not change significantly, there is a noticeable change in SAIDI and ENS values. An improvement of 20% is observed in the values of ENS from case 1 to case 4. Furthermore, the improvement is expected to be higher for larger systems. The results for each of the load points are shown in Table V. Table VI shows the system indices for all four cases from which it should be inferred that the overall system reliability has improved. The decrease in unavailability and failure rate for the highest priority load (LP9) is significant as compared to the other load points. Fig. 5 shows pictorially the improvement in the reliability at LP9. The results also indicate that the reliability indices for the most non-sensitive load does not change significantly.

## C. Sensitivity Analysis

The successful islanded operation of the microgrid in the case of upstream failure depends upon successful operation of the isolating switch. This isolation operation usually has a high probability of success. In previous scenarios, this probability was taken to be unity i.e. the islanding operation is always successful. A sensitivity analysis is performed to observe the effects of the probability of switching on system indices. Scenario 3 is considered again and the probability of successful isolation is varied from 100% to 0%. The results for all four cases are shown in Table VI. As expected, the results indicate that the decrease in probability of successfully islanding, reduces the reliability of the system. There is a decrease of 9.72% in the value of SAIDI if probability of successful islanding decreases from 1 to 0.

TABLE V
RELIABILITY INDICES FOR ALL LOAD POINTS

| LP | Index | Case 1 | Case 2 | Case 3 | Case 4 |
|---|---|---|---|---|---|
| **LP2** | λ (f/yr) | 0.726 | 0.720 | 0.721 | 0.715 |
| | r (hr) | 11.042 | 11.051 | 11.049 | 11.059 |
| | U (hr/yr) | 8.017 | 7.957 | 7.975 | 7.908 |
| **LP3** | λ (f/yr) | 0.726 | 0.669 | 0.657 | 0.635 |
| | r (hr) | 10.823 | 10.893 | 10.906 | 10.941 |
| | U (hr/yr) | 7.858 | 7.293 | 7.135 | 6.954 |
| **LP4** | λ (f/yr) | 0.726 | 0.696 | 0.706 | 0.692 |
| | r (hr) | 10.093 | 10.098 | 10.096 | 10.098 |
| | U (hr/yr) | 7.328 | 7.035 | 7.135 | 6.990 |
| **LP9** | λ (f/yr) | 0.656 | 0.248 | 0.247 | 0.291 |
| | r (hr) | 10.554 | 11.468 | 11.469 | 11.251 |
| | U (hr/yr) | 6.924 | 2.844 | 2.841 | 3.274 |

TABLE VI
RELIABILITY INDICES FOR THE MICROGRID

| Case | SAIFI (int/yr cust) | SAIDI (hr/ yr cust.) | CAIDI (hr/int cust) | ENS (kWh/yr) | AENS (kWh/yr cust.) |
|---|---|---|---|---|---|
| Case 1 | 0.721 | 7.624 | 10.57 | 42381 | 60.544 |
| Case 2 | 0.656 | 6.979 | 10.63 | 38304 | 54.714 |
| Case 3 | 0.653 | 6.967 | 10.63 | 37965 | 54.220 |
| Case 4 | 0.642 | 6.844 | 10.65 | 33821 | 48.315 |

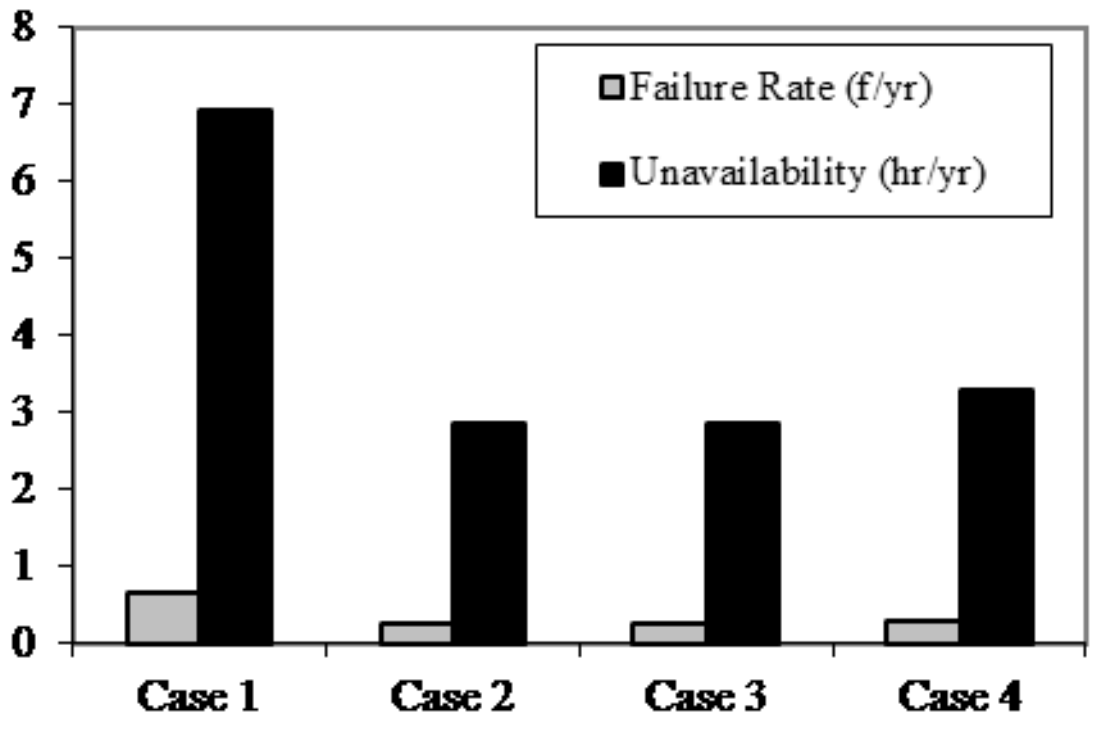

Fig. 5. Reliability indices for load points 9 for different cases.

## VI. CONCLUSION

In this work, reliability evaluation of a microgrid containing wind and solar energy sources was performed. The stochastic nature of wind and solar energy introduces complexity in the reliability evaluation methods. This stochastic nature is dealt through state-sampling simulation whereas, analytical methods are applied to evaluate the reliability metrics. The priority order of the loads is also taken into account through simulation. The studies indicated that integration of intermittent wind and solar energy sources in microgrid increase the reliability indices of the system. In particular, the most sensitive load has the largest increase in its reliability. It was also highlighted that the reliability of the system decreases with decrease in reliability of the islanding operation. For future studies, dispatchable DGs and ESSs can be included in the system to understand their effects on the reliability indices of the system. ESSs in particular, can affect the reliability of the system as they can store excess renewable energy during low-demand periods and can supply during high-demand periods. For a more accurate evaluation, hourly variations of RES generation can be utilized. Moreover, reliability cost/worth analysis can be performed.

TABLE VII
SENSITIVITY ANALYSIS

| Probability of islanding success | SAIFI (int/yr cust) | SAIDI (hr/ yr cust.) | CAIDI (hr/int cust) | ENS (kWh/yr) | AENS (kWh/ yr cust.) |
|---|---|---|---|---|---|
| 1 | 0.653 | 6.949 | 10.63 | 37960 | 54.22 |
| 0.75 | 0.670 | 7.118 | 10.62 | 39062 | 55.80 |
| 0.5 | 0.687 | 7.286 | 10.60 | 40164 | 57.37 |
| 0.25 | 0.704 | 7.456 | 10.89 | 41273 | 58.96 |
| 0 | 0.721 | 7.625 | 10.58 | 42381 | 60.54 |

## ACKNOWLEDGMENT

This work was supported by the Natural Sciences and Engineering Research Council (NSERC) of Canada.